\documentclass[conference]{IEEEtran}
\IEEEoverridecommandlockouts

\usepackage[absolute]{textpos}
\usepackage[most]{tcolorbox}
\usepackage{cite}
\usepackage{amsmath,amssymb,amsfonts}
\usepackage{algorithmic}
\usepackage{graphicx}
\usepackage{textcomp}
\usepackage{xcolor}
\usepackage{tikzsymbols}
\usepackage{algorithm}
\usepackage{algorithmic}
\usepackage{comment}
\usepackage{listings}
\usepackage{xcolor}

\definecolor{codegray}{rgb}{0.5,0.5,0.5}
\definecolor{codegreen}{rgb}{0,0.6,0}
\definecolor{codeblue}{rgb}{0,0,1}

\lstdefinestyle{mystyle}{
    basicstyle=\ttfamily\footnotesize,
    commentstyle=\color{codegreen},
    keywordstyle=\color{codeblue},
    showstringspaces=false,
    breaklines=true,
    frame=single,
    captionpos=b,
    tabsize=2,
    numbers=none,
    literate={**}{{\textbf{}}{}}2  
}
\def\BibTeX{{\rm B\kern-.05em{\sc i\kern-.025em b}\kern-.08em
    T\kern-.1667em\lower.7ex\hbox{E}\kern-.125emX}}
\begin{document}
\title{ {\fontsize{13}{13}\selectfont \textbf{This paper will be presented at IEEE VLSI Test Symposium (VTS) 2025}} \\[0.5em] ThreatLens: LLM-guided Threat Modeling and Test Plan Generation for Hardware Security Verification\\
\thanks{We thank to U.S. National Science Foundation (NSF) for their support through CAREER
Award under Grant 2339971.}
}

\author{\IEEEauthorblockN{Dipayan Saha, Hasan Al Shaikh, Shams Tarek and  Farimah Farahmandi}

\IEEEauthorblockA{\textit{Department of Electrical and Computer Engineering, University of Florida, Gainesville, FL, USA}\\
\{dsaha, hasanalshaikh, shams.tarek\}@ufl.edu, farimah@ece.ufl.edu}}
\IEEEoverridecommandlockouts
\maketitle

\begin{abstract}
Current hardware security verification processes predominantly rely on manual threat modeling and test plan generation, which are labor-intensive, error-prone, and struggle to scale with increasing design complexity and evolving attack methodologies. To address these challenges, we propose \textit{ThreatLens}, an LLM-driven multi-agent framework that automates security threat modeling and test plan generation for hardware security verification. \textit{ThreatLens} integrates retrieval-augmented generation (RAG) to extract relevant security knowledge, LLM-powered reasoning for threat assessment, and interactive user feedback to ensure the generation of practical test plans. By automating these processes, the framework reduces the manual verification effort, enhances coverage, and ensures a structured, adaptable approach to security verification. We evaluated our framework on the NEORV32 SoC, demonstrating its capability to automate security verification through structured test plans and validating its effectiveness in real-world scenarios.
\end{abstract}


\section{Introduction}

Hardware security has become increasingly critical due to the rising reliance on third-party intellectual property (IP) imports and stringent time-to-market demands. Modern system-on-chip (SoC) designs frequently incorporate reusable IP cores to minimize development costs and meet aggressive deadlines, unintentionally increasing susceptibility to hidden vulnerabilities such as hardware Trojans and backdoors \cite{mths,Bhunia2014HardwareTA}. Consequently, rigorous security verification has become essential to identify and mitigate these vulnerabilities early in the design cycle, significantly reducing post-silicon fixes and associated costs \cite{benchmark}.

Currently, a substantial portion of the verification process involves manual threat modeling and the generation of security test plans, requiring extensive domain expertise and time investment. This manual approach is inherently prone to errors, often resulting in incomplete or overlooked vulnerabilities, particularly as designs grow increasingly complex [4]. Manual verification struggles to scale efficiently, especially given rapid changes in attack methodologies and the variety of potential vulnerabilities. The limitations inherent in manual methods have underscored the urgency of automation in hardware security verification.

Large Language Models (LLMs), due to their strong natural language understanding, advanced reasoning, and knowledge transfer capabilities, present a promising avenue for automating hardware security verification processes \cite{saha2024llm}. LLMs can process design specifications or hardware description languages, effectively identifying vulnerabilities \cite{host_paper,saha2024llm} and generating related security assertions or policies \cite{asserto,meng2024nspg, kande2024security}. The hardware industry has already begun to use LLMs to automate the generation of test plans, highlighting the growing acceptance and practical benefits of these methods in real-world applications \cite{nvidia}.

However, current applications of LLMs in hardware security remain limited. For example, recent approaches successfully automate the creation and identification of security policies in RISC-V SoCs, but do not directly automate the generation of comprehensive threat models or security test plans \cite{socurellm}. Similarly, other studies effectively generate SystemVerilog assertions targeting known common weakness enumeration (CWE) vulnerabilities but do not include broader threat modeling or test planning \cite{sva}. Thus, there remains a critical gap in automating the end-to-end security verification process.


In this paper, we introduce \textit{ThreatLens}, an LLM-based agentic framework that automates security threat modeling and test plan generation, reducing manual verification efforts while ensuring robust and adaptable hardware security verification. The specific contributions of this work are multifold:
\begin{enumerate}
\item Usage of a multi-agent approach for automated security threat modeling and test plan generation through interactive engagement with verification engineers and RAG-based information retrieval.
\item Introduction of an LLM-based agent that extracts security policies from design specifications and ISA documents, effectively identifying software-exploitable hardware vulnerabilities.
\item Demonstration of the efficacy of ThreatLens on the Neorv32 SoC, validating its ability to automate security verification through real-world SoC designs.
\end{enumerate}

\section{Proposed Framework}
\label{sec:proposed_framework}
\begin{figure}[t]
\centering
\includegraphics[scale=.4]{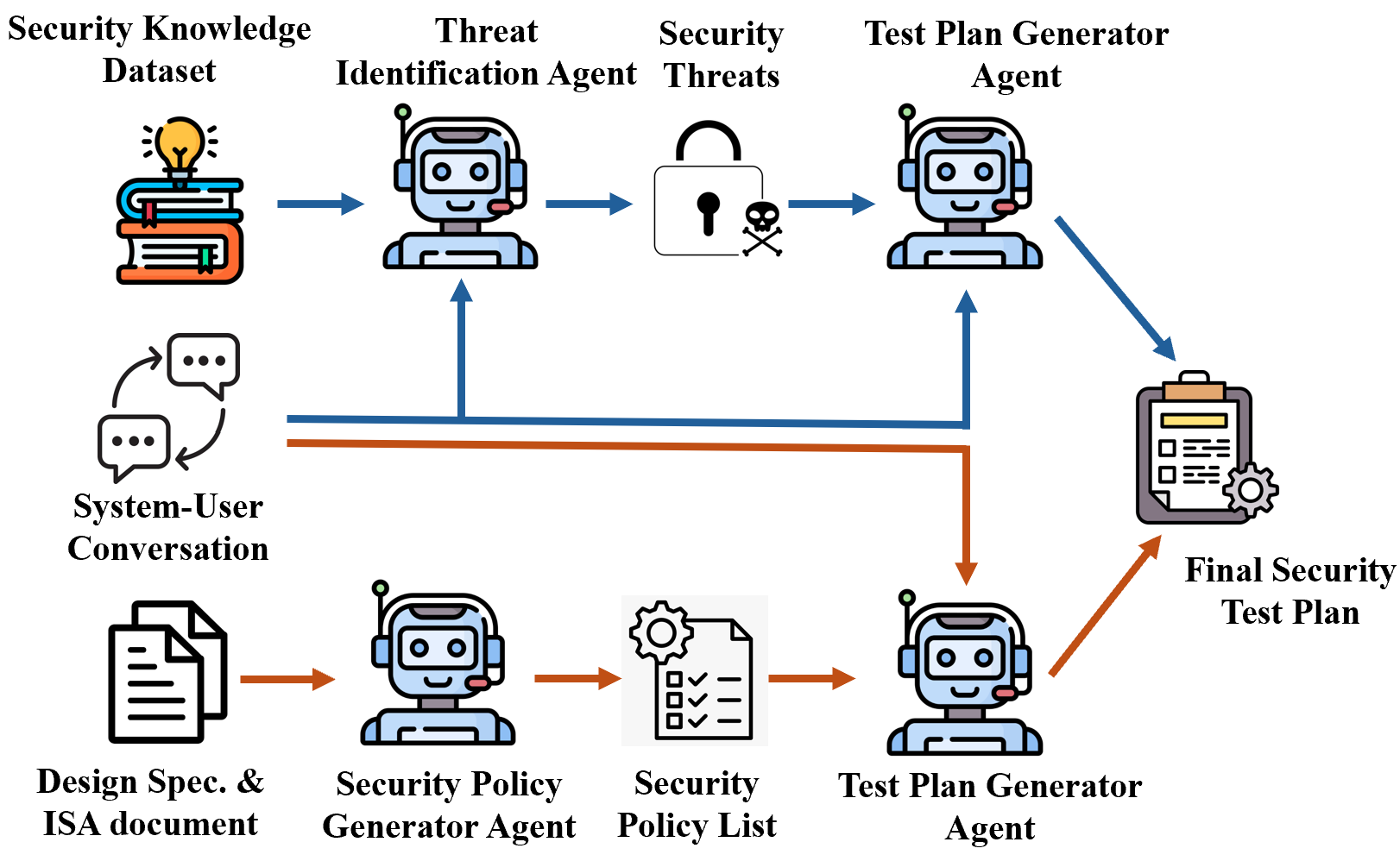}
\caption{Overview of proposed \textit{ThreatLens} framework. The blue arrows indicate the flow for test plan generation for physical and supply chain attacks and the orange arrows indicate the flow for that of software-exploitable hardware vulnerabilities.}
\label{fig:proposed_framework}
\vspace{-0.15in}
\end{figure}

The \textit{ThreatLens} framework, shown in Figure \ref{fig:proposed_framework}, is a multi-agent system designed for hardware security threat modeling and test plan generation. As an agentic-based approach, \textit{ThreatLens} consists of four specialized agents, each responsible for a specific task. Through an iterative system-user conversation, the framework gathers design details, security assumptions, verification plans, and testing capabilities from the verification engineer. \textit{ThreatLens} framework operates through two distinct flows:
\begin{itemize}
\item \textit{\underline{Flow 1}}: It deals with physical and supply chain attacks. In this flow, the \textit{Threat Identification Agent} retrieves security threat models from a security knowledge dataset, filtering and prioritizing threats based on user input. The \textit{Test Plan Generator Agent} then formulates a hardware security test plan to assess risks such as side-channel attacks, fault injection, and reverse engineering.
\item \textit{\underline{Flow 2}}: It is related software-exploitable vulnerabilities. In this flow, the \textit{Security Policy Generator Agent} extracts security policies from the Instruction Set Manual and design specification document, generating a Security Policy List. This information is processed by the \textit{Test Plan Generator Agent}, which develops a verification strategy to assess vulnerabilities such as privilege escalation, access control violations, microarchitectural side channels, and memory corruption.
\end{itemize}


\subsection{Threat Identification Agent}
\label{sec:threat_identification_agent}
\begin{figure}[t]
\centering
\includegraphics[scale=.4]{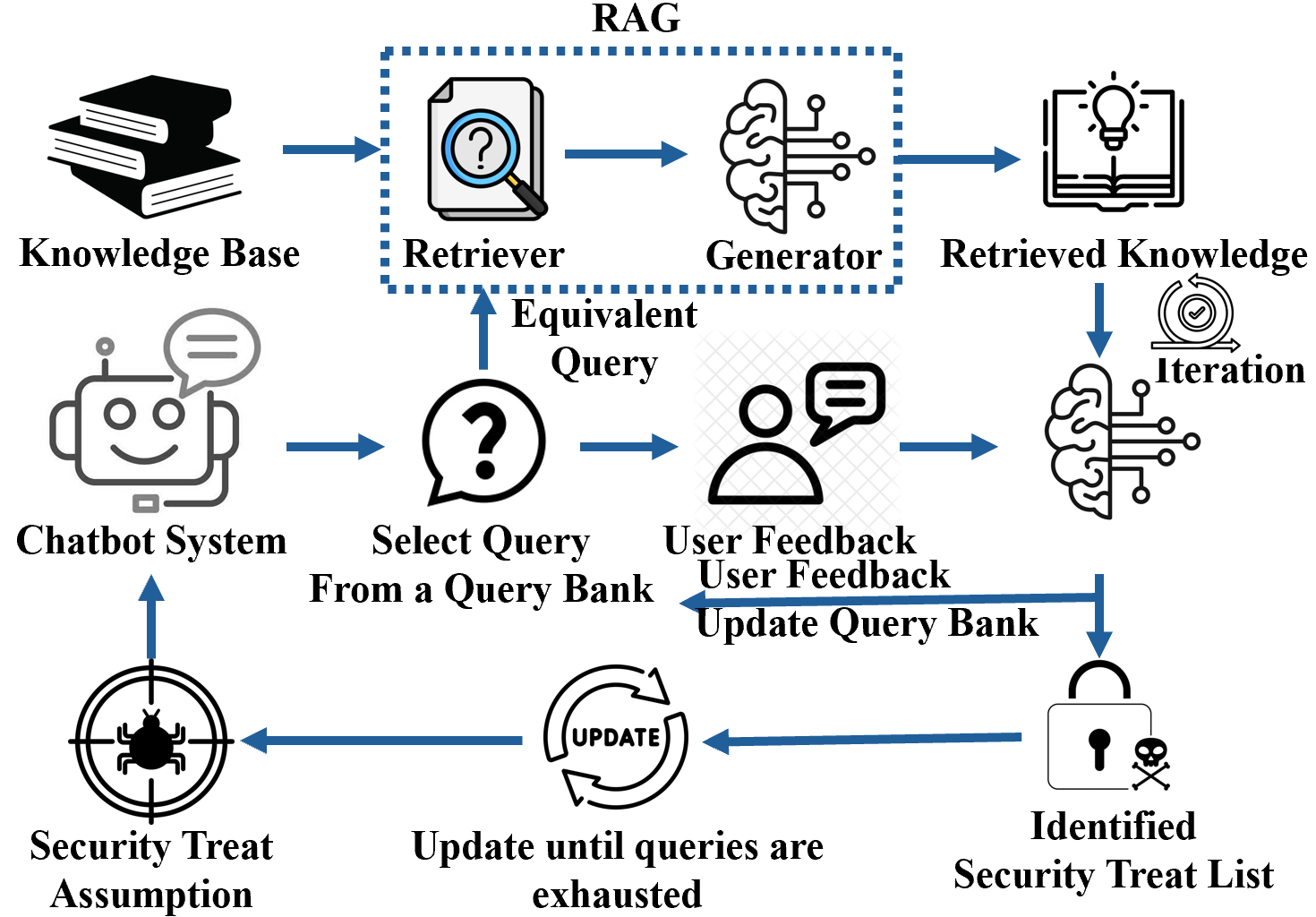}
\caption{Overview of Threat Identification Agent.}
\label{fig:threat_identification_agent}
\vspace{-0.15in}
\end{figure}

 The \textit{Threat Identification Agent} operates in a multistep, iterative manner, combining RAG-based knowledge retrieval, LLM-driven reasoning, and interactive user feedback. As shown in Figure \ref{fig:threat_identification_agent}, this process is designed to systematically extract relevant security knowledge, engage with the verification engineer, assess the relevance of potential threats using an LLM, and refine the security threat list iteratively.

\textit{\underline{Step 1: Security Knowledge Extraction}}:
The process begins with the extraction of security knowledge, where the agent focuses on predefined categories of physical attack methodologies, including side-channel attacks, fault injection techniques, reverse engineering, IC cloning, and invasive hardware attacks. Since these threats require highly specific security insights, the framework maintains a knowledge base of academic papers that document state-of-the-art attack models. To ensure the retrieval of accurate and contextually relevant information, the framework employs a RAG-based methodology. A predefined set of queries is used to systematically extract security knowledge corresponding to each of the threats under consideration. This retrieved information serves as a critical foundation for evaluating the relevance of security threats in the subsequent stages.

\textit{\underline{Step 2: Initial Query Generation and User Feedback}}:
Once the security knowledge is extracted, the process transitions to this step. A Chatbot System engages with the verification engineer to collect relevant details about the system under evaluation, including design implementation specifics, application context, supply chain risks, and security assumptions. These queries are predefined within the system and are strategically structured to capture essential security aspects of the design. At the beginning of the process, the system presents the engineer with an initial query and records their response. Feedback received in this step ensures that the subsequent threat evaluation process is grounded in the actual security requirements and constraints of the system.

\textit{\underline{Step 3: LLM-Based Security Threat Identification}}:
In this step, the verification engineer’s response is included in a structured LLM prompt along with the security knowledge retrieved in the previous step. The system then formulates individual prompts for each security threat from the predefined list and conducts LLM-based inference to evaluate their relevance. Using its reasoning capability, the LLM generates an updated threat list that retains only those threats deemed relevant to the given system. This step serves as the core decision-making process within the \textit{Threat Identification Agent}, as it dynamically filters out threats that do not align with the design context.

\textit{\underline{Step 4: Query Refinement and Threat List Update}}:
To further refine the threat list, this step is introduced as an iterative process. The system continues engaging with the verification engineer, presenting additional queries, and incorporating new responses into the threat assessment framework. Since many security queries are interconnected, the query bank is dynamically updated, allowing for the removal of redundant or now-irrelevant queries based on prior user feedback. The threat list is also continuously refined in each iteration, ensuring that only genuine, design-specific threats remain under consideration. This process continues until all predefined queries have been exhausted and no further refinements are required. In this way, after the execution of these four steps, a finalized list of security threats is generated.

\subsection{Security Policy Generator Agent}
\label{sec:security_policy_generator_agent}
The \textit{Security Policy Generator Agent}, shown in Figure \ref{fig:security_policy_generator_agent} operates within the software-exploitable hardware vulnerability modeling workflow, focusing on threats such as privilege escalation, memory corruption, access control violations, and other software-exploitable hardware vulnerabilities. This agent is responsible for extracting design-specific security policies from user-provided specification files and instruction set architectures (ISA) using RAG-based knowledge retrieval. The extracted policies serve as the foundation for identifying security vulnerabilities and formulating a test plan for verification. The process consists of the following key steps:

\begin{figure}[t]
\centering
\includegraphics[scale=.45]{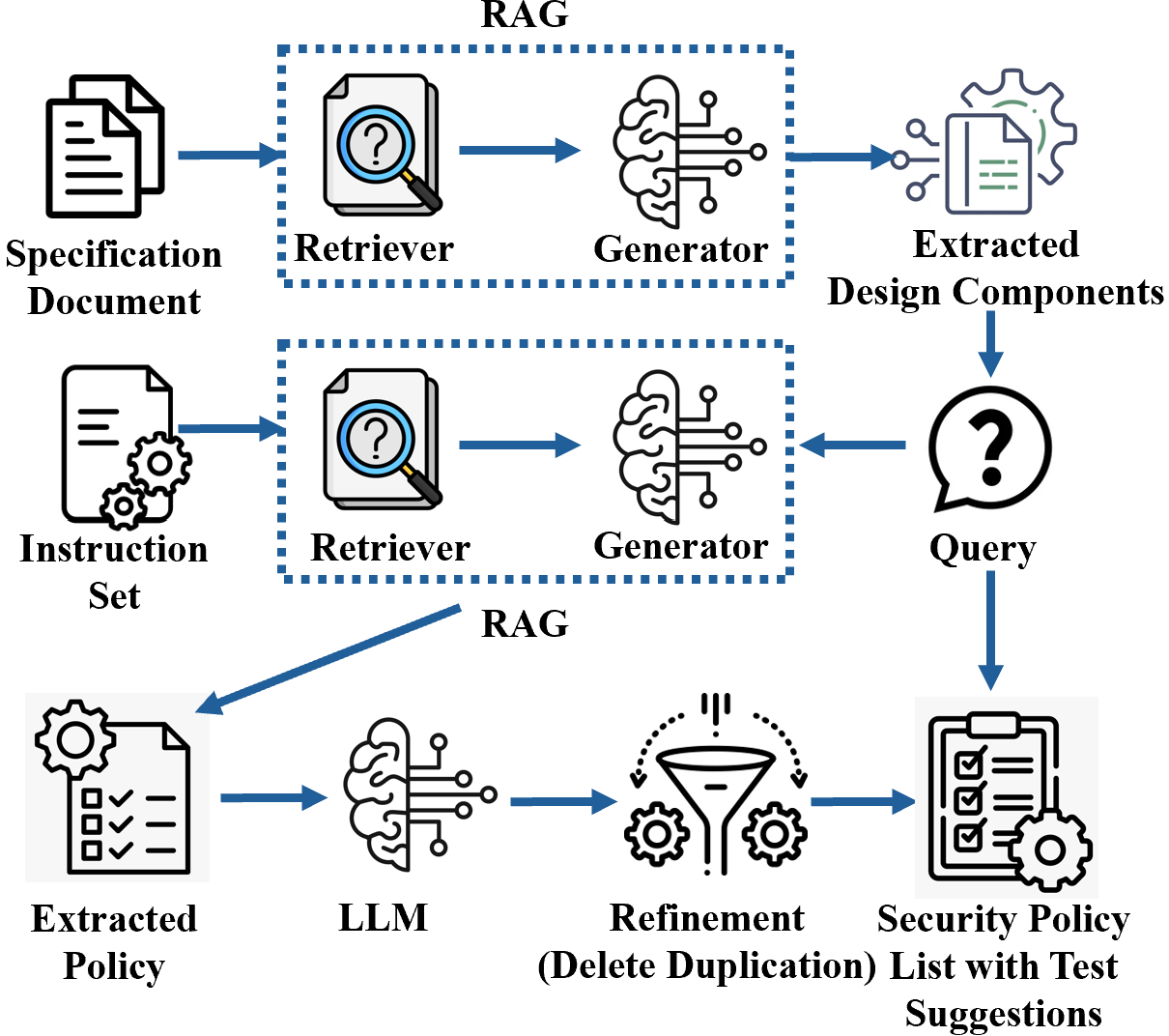}
\caption{Overview of Security Policy Generator Agent.}
\label{fig:security_policy_generator_agent}
\vspace{-0.15in}
\end{figure}

\textit{\underline{Step 1: Extraction of Design-Specific Components}}: The process begins by extracting design-specific information from a user-provided specification document. Since a specification document is too long to be analyzed through LLM in a single inference, we used the RAG system for this purpose. The system extracts registers and instructions used in the design through a retriever and an LLM used as a generator. Later, the list is refined by removing duplicate mentions from the extracted list.

\textit{\underline{Step 2:  Extraction of Policies from ISA}}: To ensure comprehensive coverage of security vulnerabilities, in this step, the agent retrieves relevant security policies from the ISA document. A second RAG system is employed to extract policy information from ISA documentation. The RAG searches the ISA for policies related to every element of the list extracted in Step 1.

\textit{\underline{Step 3: Security Policy and Threat Identification}}: 
Once policies are extracted, they are passed to an LLM. The LLM analyzes the extracted policies in the context of the mentioned software-exploitable hardware security threats. Guided by a prompt with specific instructions, the LLM separates the security policies, mentions their security relevance, and also highlights possible security risks such as privilege escalation, access control weaknesses, and memory corruption. 

\subsection{Test Plan Generator Agent}
\label{sec:test_plan_generator_agent}
\begin{figure}[t]
\centering
\includegraphics[scale=.50]{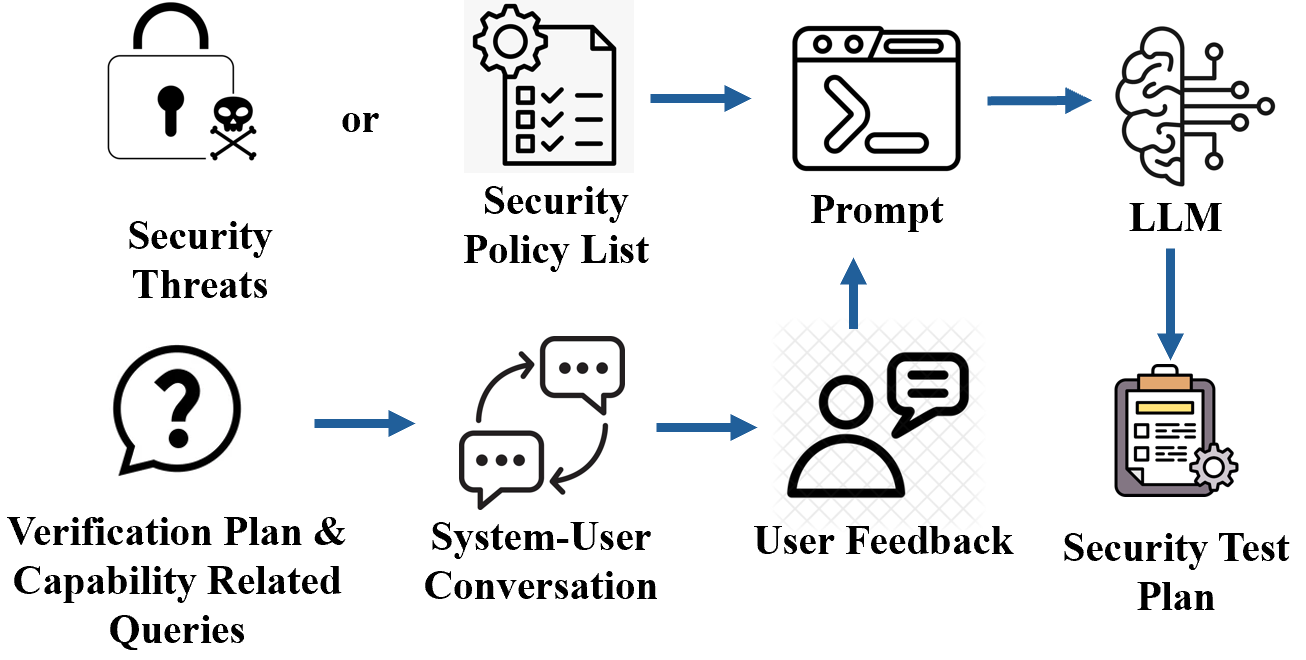}
\caption{Overview of Test Plan Generator Agent.}
\label{fig:test_plan_generator_agent}
\vspace{-0.15in}
\end{figure}

The \textit{Test Plan Generator Agent}, illustrated in Figure \ref{fig:test_plan_generator_agent}, plays a crucial role in both the modeling of physical threats and the analysis of vulnerabilities exploitable by software. Although the core workflow remains the same across both flows, the inputs to the agent differ depending on the type of security threat being analyzed. In the physical \& supply chain security verification workflow, the agent receives a list of security threats identified in Section \ref{sec:threat_identification_agent}. In contrast, within the software-exploitable hardware vulnerability analysis workflow, the input consists of a security policy list extracted in Section \ref{sec:security_policy_generator_agent}. 

The \textit{Test Plan Generator Agent} initiates a structured system-user conversation to determine the feasibility of verification. During this interaction, the system presents a set of pre-defined queries to the verification engineer, gathering information about the available security testing infrastructure, verification tools, budget, and time allocation. After gathering information about the available verification capabilities, the system constructs a structured prompt that combines security threat data or policy enforcement requirements with the verification constraints provided by the engineer. This prompt also includes specific information about verification methodologies based on existing industry practices. The generated security test plan consists of multiple test cases in which each test case includes information about the threat, the test objective, the test methodology, the expected result, the evaluation criteria, and the test tool.

\begin{figure}[!t]
\begin{tcolorbox}[colback=blue!10!white, colframe=gray!50!blue,  fontupper=\small, title=\textbf{Listing I: An Abridged Form of a Sample Test Case from Generated from Test Plan.}]
\textbf{Threat Category:} Unauthorized Access, Memory Access
\newline \textbf{Test Objective:} The test aims to verify that the execution of a load or load-reserved instruction which accesses a physical address within a region without read permissions, raises a load access-fault exception...
\newline \textbf{Test Methodology:}
\newline - \textit{Formal Verification:} ... In this case, formal methods are to be used to ensure that memory read access controls appropriately raise a load access-fault exception when violated.
\newline - \textit{Emulation:} ... The emulated hardware should raise a load access-fault exception when a read access violation is attempted, under the same conditions as specified in the security policy.
\newline - \textit{Simulation:} Create a software model of the chip design for verification. The simulation should prompt a load access-fault exception when a violation of the read access control is simulated.
\newline \textbf{Expected Result:}
\newline- \textit{Formal Verification:} The verification proves irrespective of the input conditions, an access-fault is raised when an illegal read operation is attempted within a protected memory region.
\newline- \textit{Emulation:} The emulated hardware raises an access-fault when an illegal read is attempted within a protected memory area, replicating the live environment successfully.
\newline- \textit{Simulation:} The software model correctly raises a load access-fault exception when the read access control is violated.

\textbf{Evaluation Criteria:}
\newline- \textit{Formal Verification:} Formal verification fails if it does not categorically prove that an access-fault exception is raised whenever the specified condition is violated.
\newline- \textit{Emulation:} Evaluation is unsuccessful if the emulated hardware does not raise an access-fault exception under the designated violation condition.
\newline- \textit{Simulation:} The test is deemed to fail if the software model does not behave as expected by the security policy when a read access violation is attempted.

\textbf{Testing Tool:}
\newline- \textit{Formal Verification:} JasperGold
\newline- \textit{Emulation:} Zebu
\newline- \textit{Simulation:} Modelsim, VCS
\end{tcolorbox}
\vspace{-0.15in}
\end{figure}

\section{Experimental Results}
\label{sec: results}
To evaluate the suitability of our proposed method, we have selected the Neorv32 SoC \cite{neorv32} as the experimental testing platform. This platform features a compact, customizable, and extensible MCU-class 32-bit RISC-V microcontroller-like SoC. In this experiment, we used GPT-4o as the LLM in the proposed \textit{ThreatLens} framework. RAG systems used in the framework utilize the LangChain platform, OpenAI embeddings for semantic representation, and Facebook AI similarity search (FAISS) \cite{faiss} as the search algorithm for efficient similarity-based document retrieval. An abridged version of a generated test case is shown in Listing I.

\textit{ThreatLens} systematically developed a comprehensive set of security policies through a detailed analysis of both the specifications of the devices and the documents pertaining to their ISA. For the NEORV32 system, the framework automatically generated 854 unique security policies, a task that would require significant time and effort if performed manually. To demonstrate the relevance, practical utility, and reliability of these outcomes, we provide an analysis of two distinct case studies, each of which is focused on a specific policy output of \textit{ThreatLens}.

\subsubsection{Case Study 1}\label{subsubsec: gen_SP1}
ThreatLens generated the following policy statement by analyzing the ISA documents: ``attempting to execute a load or load-reserved instruction which accesses a physical address within a PMP region without read permissions raises a load access-fault exception." This policy, derived from the privileged ISA document of the RISC-V ISA, addresses the security requirements related to physical memory protection (PMP).

PMP is a hardware-based security feature in RISC-V systems designed to manage access control across varying software privilege levels to memory spaces. 
Utilizing this feature ensures that lower-privileged software is unable to read (i.e., load) privileged memory.
Previous research in the hardware security domain confirms the necessity of complying with this policy. 
For instance, researchers and industry professionals studied a vulnerability that reflects non-compliance with this policy during the well-known hardware hackathon HACK@DAC\cite{li2020conference}. 
Moreover, as detailed in \cite{al2023sharpen}, an open-source RISC-V SoC was reported in the real world, where the DMA allowed lower privileged software to circumvent PMP protections and load sensitive data from privileged memory. 

\subsubsection{Case Study 2} 
\label{subsubsec: gen_SP3}

\textit{ThreatLens} also suggested that the policy ``attempts to access a non-existent CSR raise an illegal instruction exception" should be verified as it can lead to ``unauthorized access" and potentially cause ``integrity, availability" violations. We note that a vulnerability that results in the violation of this policy has been reported to be found in open-source RISC-V processors by authors in \cite{gohil2024mabfuzz, chen2023psofuzz}. 

These case studies demonstrate that failing to verify the policies generated by \textit{ThreatLens} can result in significant real-world security implications, as such violations have been observed in practice. 
Consequently, \textit{ThreatLens} can serve as a highly valuable tool for verification engineers, aiding in the identification of system security requirements (i.e., policies), associated vulnerabilities, and potential threats, thereby facilitating the automatic formulation of an effective test plan.

\section{Conclusion}
\label{sec:conclusion}
\textit{ThreatLens} paves the way for future advancements in LLM-driven security verification methodologies. Although it effectively identifies security threats and generates structured test plans, it does not incorporate security asset extraction, which could optimize the threat modeling process by reducing redundant queries and improving efficiency. Furthermore, our implementation relies on GPT-4o for inference, which, while powerful, is a closed-source model with high computational costs. Future work will focus on integrating security asset extraction to streamline threat identification, exploring open-source LLMs for cost-effective inference, and enhancing adaptability to diverse hardware architectures, ensuring broader applicability and improved efficiency in security verification.
\bibliographystyle{IEEEtran}
\bibliography{IEEEabrv,eprint_version}

\begin{thebibliography}{10}
\providecommand{\url}[1]{#1}
\csname url@samestyle\endcsname
\providecommand{\newblock}{\relax}
\providecommand{\bibinfo}[2]{#2}
\providecommand{\BIBentrySTDinterwordspacing}{\spaceskip=0pt\relax}
\providecommand{\BIBentryALTinterwordstretchfactor}{4}
\providecommand{\BIBentryALTinterwordspacing}{\spaceskip=\fontdimen2\font plus
\BIBentryALTinterwordstretchfactor\fontdimen3\font minus \fontdimen4\font\relax}
\providecommand{\BIBforeignlanguage}[2]{{%
\expandafter\ifx\csname l@#1\endcsname\relax
\typeout{** WARNING: IEEEtran.bst: No hyphenation pattern has been}%
\typeout{** loaded for the language `#1'. Using the pattern for}%
\typeout{** the default language instead.}%
\else
\language=\csname l@#1\endcsname
\fi
#2}}
\providecommand{\BIBdecl}{\relax}
\BIBdecl

\bibitem{mths}
M.~Tehranipoor, H.~Salmani, and X.~Zhang, \emph{Integrated Circuit Authentication: Hardware Trojans and Counterfeit Detection}, 04 2014.

\bibitem{Bhunia2014HardwareTA}
\BIBentryALTinterwordspacing
S.~Bhunia, M.~S. Hsiao, M.~Banga, and S.~Narasimhan, ``Hardware trojan attacks: Threat analysis and countermeasures paper is a survey of the state-of-the-art trojan attacks, modeling,,'' 2014. [Online]. Available: \url{https://api.semanticscholar.org/CorpusID:110420508}
\BIBentrySTDinterwordspacing

\bibitem{benchmark}
S.~Tarek, H.~A. Shaikh, S.~R. Rajendran, and F.~Farahmandi, ``Benchmarking of soc-level hardware vulnerabilities: A complete walkthrough,'' in \emph{2023 IEEE Computer Society Annual Symposium on VLSI (ISVLSI)}, 2023, pp. 1--6.

\bibitem{saha2024llm}
D.~Saha, S.~Tarek, K.~Yahyaei, S.~K. Saha, J.~Zhou, M.~Tehranipoor, and F.~Farahmandi, ``Llm for soc security: A paradigm shift,'' \emph{IEEE Access}, 2024.

\bibitem{host_paper}
D.~Saha, K.~Yahyaei, S.~Kumar~Saha, M.~Tehranipoor, and F.~Farahmandi, ``Empowering hardware security with llm: The development of a vulnerable hardware database,'' in \emph{2024 IEEE International Symposium on Hardware Oriented Security and Trust (HOST)}, 2024, pp. 233--243.

\bibitem{asserto}
S.~S. Miftah, A.~Srivastava, H.~Kim, and K.~Basu, ``Assert-o: Context-based assertion optimization using llms,'' in \emph{Proceedings of the Great Lakes Symposium on VLSI 2024}, 2024, pp. 233--239.

\bibitem{meng2024nspg}
X.~Meng, A.~Srivastava, A.~Arunachalam, A.~Ray, P.~H. Silva, R.~Psiakis, Y.~Makris, and K.~Basu, ``Nspg: Natural language processing-based security property generator for hardware security assurance,'' in \emph{Proceedings of the 61st ACM/IEEE Design Automation Conference}, 2024, pp. 1--6.

\bibitem{kande2024security}
R.~Kande, H.~Pearce, B.~Tan, B.~Dolan-Gavitt, S.~Thakur, R.~Karri, and J.~Rajendran, ``(security) assertions by large language models,'' \emph{IEEE Transactions on Information Forensics and Security}, 2024.

\bibitem{nvidia}
\BIBentryALTinterwordspacing
Nvidia developer. [Online]. Available: \url{https://developer.nvidia.com/blog/building-ai-agents-to-automate-software-test-case-creation/}
\BIBentrySTDinterwordspacing

\bibitem{socurellm}
\BIBentryALTinterwordspacing
S.~Tarek, D.~Saha, S.~K. Saha, M.~Tehranipoor, and F.~Farahmandi, ``{SoCureLLM}: An {LLM}-driven approach for large-scale system-on-chip security verification and policy generation,'' Cryptology {ePrint} Archive, Paper 2024/983, 2024. [Online]. Available: \url{https://eprint.iacr.org/2024/983}
\BIBentrySTDinterwordspacing

\bibitem{sva}
\BIBentryALTinterwordspacing
Q.~Zhang, W.~Sun, C.~Fang, B.~Yu, H.~Li, M.~Yan, J.~Zhou, and Z.~Chen, ``Exploring automated assertion generation via large language models,'' \emph{ACM Trans. Softw. Eng. Methodol.}, vol.~34, no.~3, Feb. 2025. [Online]. Available: \url{https://doi.org/10.1145/3699598}
\BIBentrySTDinterwordspacing

\bibitem{neorv32}
\BIBentryALTinterwordspacing
{stnolting}. (2024) The neorv32 soc. [Online]. Available: \url{https://github.com/stnolting/neorv32}
\BIBentrySTDinterwordspacing

\bibitem{faiss}
M.~Douze, A.~Guzhva, C.~Deng, J.~Johnson, G.~Szilvasy, P.-E. Mazaré, M.~Lomeli, L.~Hosseini, and H.~Jégou, ``The faiss library,'' 2024.

\bibitem{li2020conference}
Z.~Li, ``Conference report from the 57th design automation conference,'' \emph{IEEE Design \& Test}, vol.~37, no.~6, pp. 99--101, 2020.

\bibitem{al2023sharpen}
H.~Al-Shaikh, A.~Vafaei, M.~M.~M. Rahman, K.~Z. Azar, F.~Rahman, F.~Farahmandi, and M.~Tehranipoor, ``Sharpen: Soc security verification by hardware penetration test,'' in \emph{Proceedings of the 28th Asia and South Pacific Design Automation Conference}, 2023, pp. 579--584.

\bibitem{gohil2024mabfuzz}
V.~Gohil, R.~Kande, C.~Chen, A.-R. Sadeghi, and J.~Rajendran, ``Mabfuzz: Multi-armed bandit algorithms for fuzzing processors,'' in \emph{2024 Design, Automation \& Test in Europe Conference \& Exhibition (DATE)}.\hskip 1em plus 0.5em minus 0.4em\relax IEEE, 2024, pp. 1--6.

\bibitem{chen2023psofuzz}
C.~Chen, V.~Gohil, R.~Kande, A.-R. Sadeghi, and J.~Rajendran, ``Psofuzz: Fuzzing processors with particle swarm optimization,'' in \emph{2023 IEEE/ACM International Conference on Computer Aided Design (ICCAD)}.\hskip 1em plus 0.5em minus 0.4em\relax IEEE, 2023, pp. 1--9.

\end{thebibliography}

\end{document}